\documentclass[letter]{aa} % for the letters 
\usepackage{graphicx}
\usepackage{natbib}
\usepackage{txfonts}
\usepackage{textcomp}
\bibpunct{(}{)}{;}{a}{}{,}             %% natbib format for A&A and ApJ
\begin{document} 

   \title{Reinforcing the link between the double red clump and the X-shaped bulge of the Milky Way\thanks{Based on observations taken within the ESO VISTA Public Survey VVV, Programme ID 179.B-2002}}

%   \subtitle{NGC 4710 observed with MUSE@VLT\footnote{Based on observations collected at the ESO La Silla-Paranal Observatory within MUSE science verification program 60.A-9307(A).}}

   \author{O. A. Gonzalez\inst{1,2}
          \and 
          M. Zoccali\inst{3,4}
          \and Victor P. Debattista\inst{5}
          \and J. Alonso-Garc\'{\i}a\inst{6, 4}
          \and E. Valenti\inst{7}
          \and D. Minniti\inst{4, 8, 9}
           }

   \institute{European Southern Observatory, Ave. Alonso de Cordova 3107, Casilla 19, 19001, Santiago, Chile \\
   \email{ogonzale@eso.org}
    \and
       Institute for Astronomy, University of Edinburgh, Royal Observatory, Blackford Hill, Edinburgh, EH9 3HJ,  UK
        \and
             Instituto de Astrof\'{\i}sica, Facultad de F\'{\i}sica, Pontificia Universidad Cat\'olica de Chile, Av. Vicu\~na Mackenna 4860, Santiago 22, Chile
     \and
        Instituto Milenio de Astrof\'{\i}sica, Santiago, Chile
        \and
        Jeremiah Horrocks Institute, University of Central Lancashire, Preston PR1 2HE, UK
        \and
        Unidad de Astronom\'{\i}a, Facultad Cs. B\'asicas, Universidad de Antofagasta, Avda. U. de Antofagasta 02800, Antofagasta, Chile
       \and 
        European Southern Observatory, Karl-Schwarzschild Strasse 2, D-85748 Garching, Germany
        \and
       Departamento de Ciencias F\'isicas, Universidad Andr\'es Bello, Rep\'ublica 220, Santiago, Chile
       \and
       Vatican Observatory, V00120 Vatican City State, Italy.  
                }

   \date{Received August 15, 2014; accepted August 16, 2014}

% \abstract{}{}{}{}{} 
% 5 {} token are mandatory
 
  \abstract{The finding of a double red clump in the  luminosity function of the Milky Way bulge has been
    interpreted as  evidence for  an X-shaped  structure. Recently, an  alternative explanation  has been
    suggested, where  the double red  clump is an  effect of multiple  stellar populations in  a classical
    spheroid. In this letter we provide an observational assessment of this scenario and show that it is not
    consistent with the behaviour of the red clump across different lines of sight, particularly at high
    distances  from the  Galactic plane.  Instead, we confirm that  the shape of the  red clump  magnitude
    distribution closely follows  the distance distribution expected for an  X-shaped bulge
    at critical Galactic latitudes. We also emphasize some key observational properties of the bulge 
    red clump that should not be neglected in the search for alternative scenarios.}

\titlerunning{Reinforcing the link of the double red clump with the X-shape of the Milky Way bulge}
\maketitle
%
%________________________________________________________________

\section{Introduction}

Understanding galaxy formation and evolution is one of the fundamental goals of modern astronomical research. In particular, modern simulations reveal that the overall shape of galaxy bulges and the properties of their stars can hold the fingerprints of the role that different processes, such as dynamical instabilities, hierarchical merging, and dissipative collapse, played in the assembly history of the entire host galaxy.

There are two main ideas in how bulges form: the merger-driven bulge scenario and the secular evolution scenario. Bulges can be formed during the early stages of the galaxy, dominated by violent and rapid processes of hierarchical merging of sub-clumps of dark-matter carrying baryons and gas \citep[e.g.][]{brook+11}. These processes give rise to a spheroidal structure that is dominated by old stars and is known in the literature as a classical bulge \citep[see][for a recent review]{brook-christensen+15}. On the other hand, the bulge structure in the secular evolution scenario is naturally born from the dynamical evolution  of  a stellar  disc  \citep[e.g.][]{ComSan81, Ath05b}. The stellar bar in the inner regions of a galaxy suffers from so-called buckling instabilities that cause it to thicken in the vertical direction. This produces the so-called boxy-peanut (B/P) bulges, which earned this name
because their shape is  boxy, like a peanut, or even like an X when seen in  edge-on projection \citep[see][for a recent review]{laurikainen+15}\footnote{We note that attempting to classify observed bulges as the result of only one of these two scenarios is perhaps a too simplistic view. Recent observations of main-sequence galaxies at z$\sim$2 show that the migration and central coalescence of massive star-forming clumps \citep{ImmSamGer04, bournaud+09} may develop into a global violent disc instability and also lead to the formation of compact bulges through dissipation \citep{dekel-burkert+14}.}. 
%main-sequence galaxies at z$\sim$2 show large rotating discs, characterised by very high-velocity dispersion ($\sim$50-100 km/s) and large-scale (several kpc) star-forming clumps \citep[e.g.][]{forster-schreiber+09, newman+13}. . In this sense, bulge formation can be a mixture of very complex processes and thus understanding the nature of the bulge of the Milky Way by investigating its resolved stellar populations becomes critical.

\begin{figure*}[ht]
\centering
\includegraphics[width=13cm,angle=0]{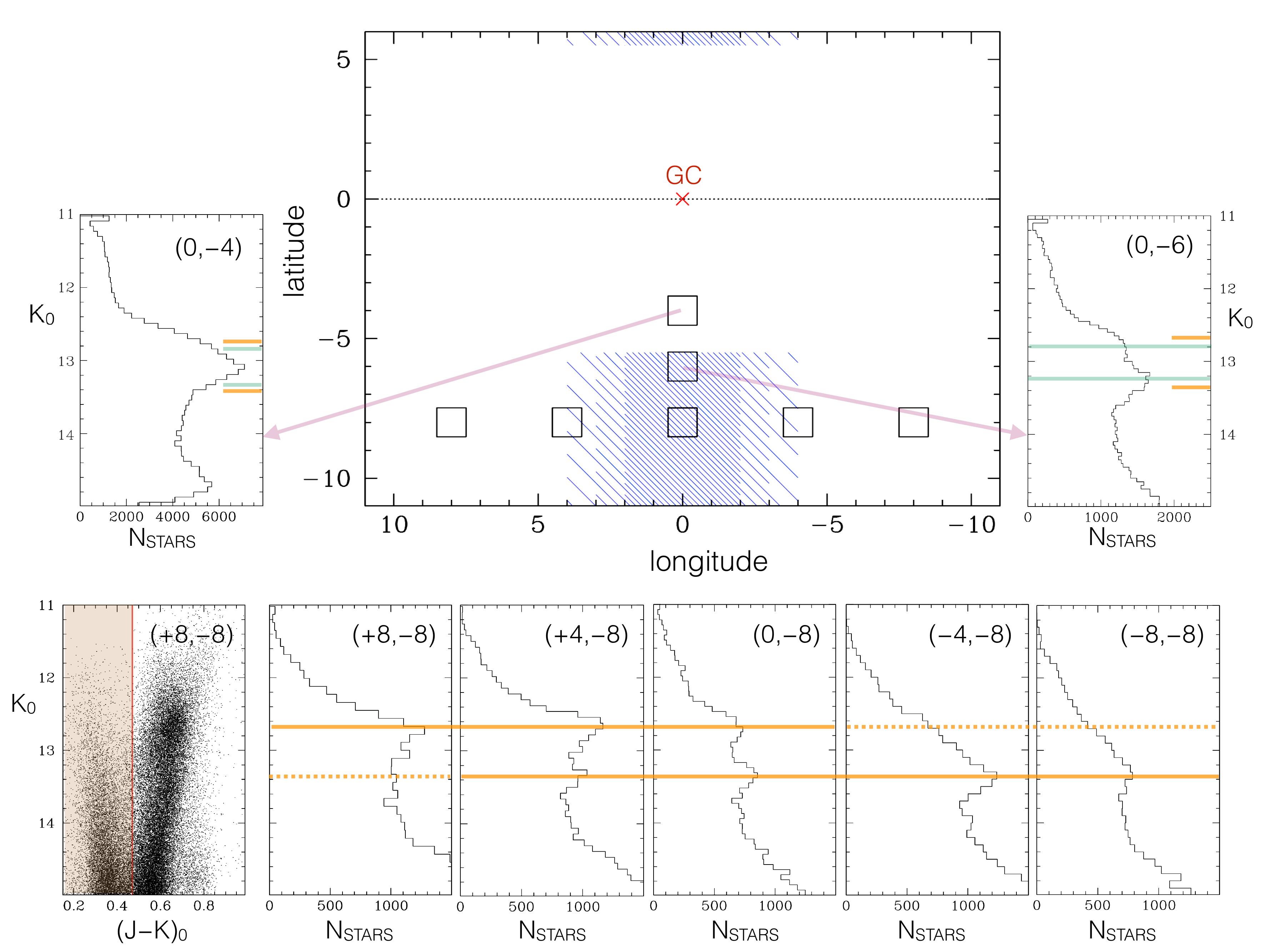}\\
\caption{Luminosity function of the bulge RGB and RC, in several fields. 
The large box in the middle shows the bulge area covered by the VVV survey, 
with the seven fields for which we show the RGB luminosity function.
We also show as a shaded blue rectangle the region where the double RC is visible
($\mathrm{|b|  >  5^{\circ}}$  $\mathrm{|l|<4^{\circ}}$). The shading
is finer where the double RC is clearly evident and becomes wider when 
one of the two RCs becomes significantly weaker than the other.
In the bottom left corner we show the CMD of the field at ($+8^\circ,-8^\circ$)
as an example of the colour selection that we applied to exclude 
the main sequence of the foreground disc. Orange horizontal lines mark
the magnitude of the two RCs in the five fields at b$=-8^\circ$ shown in
the bottom panels. Green horizontal lines mark
the magnitude of the two RCs at b$=-6^\circ$, shown at the top right. 
Only one RC is visible at b$=-4^\circ$ (Baade's Window), shown at the top left.}
\label{fig1}
\end{figure*}
%Recent results based on the interpretation of the observed metallicity distribution, kinematics and spatial distribution of bulge stars seem to suggest a complex nature \citep{gonzalez-gadotti+15}.
In response to this, attempts to link the formation of the Galactic bulge to these processes have increased considerably during the past years. In particular, the structure of the bulge has been mapped extensively thanks to the large coverage of recent near-IR surveys of the inner Galaxy. These studies are often based on the construction of the luminosity function of the bulge towards a given line of sight, where the red clump (RC) feature can be identified,  its mean magnitude measured, and used as a distance indicator \citep[e.g.][]{stanek+98}. Following the discovery of a split of the RC along the bulge's minor axis by \citet{zoccali+10} and \citet{mcwilliam-fulbright-rich+10}, \citet{mcwilliam-zoccali+10}, \citet{nataf+10}, and \citet{Saito+11} used this technique to provide a wide mapping of the double RC feature in the bulge luminosity function. These studies concluded that the bright (bRC) and faint (fRC) red clumps are the consequence of two over-densities of stars located at different distances, namely the two southern arms of an X-shaped structure that both cross the lines of sight. \citet{wegg-gerhard+13} modelled the distribution of RC stars, observed in the Vista Variables in the Via Lactea (VVV) ESO public survey and provided the complete mapping of what is now referred to as the X-shaped bulge of the Milky Way. 

Recently, \citet{lee+15} presented a different interpretation for the split RC in which a  classical bulge with an additional population enriched in helium co-exists with a bar. Within this hypothesis, the authors assigned all the bar-like properties to the Milky Way bar component, which has not undergone a buckling instability and is thus restricted to low Galactic latitudes. The double RC is then not caused by the X-shape of the bar, but instead is the consequence of a massive classical bulge with a significant fraction of stars enriched in helium. In this letter, we challenge the validity of this scenario based on the observational properties of the RC in some specific lines of sight.

%________________________________________________________________

\section{Observational requirements for interpreting the double red clump}

In this section we list four of the basic observed characteristics of the RC magnitude distribution of the Galactic bulge. Any physical interpretation of the double RC must satisfy, as a minimum requirement, the following purely observational properties:

\begin{enumerate}
\item At  Galactic  latitudes  $\mathrm{|b|  <  5^{\circ}}$,  a  single  RC  is  observed  \citep{babusiaux+05, rattenbury+07,   cao+13,   gonzalez+13}.   An   additional   much   fainter   component   has   also   been detected\footnote{This  additional  bump  seen in  the  luminosity  function  is  found at  much  fainter magnitudes ($\mathrm{K_{s_0}\sim14.3}$) and it does not affect the conclusions of this work or those from \citet{lee+15}. Its origin has been linked to the bulge red giant branch bump \citep{nataf+10} or the background stellar  disc \citep{gonzalez+11c}, however, this  discussion is beyond the  scope of this letter.}. The  mean magnitude of the  red clump varies by 1.0 mag from $\mathrm{l=+10^{\circ}}$ (where $\mathrm{K_{s_0}(RC)=12.4}$) to  $\mathrm{l=-10^{\circ}}$ (where $\mathrm{K_{s_0}(RC) =13.4}$).

\item At  Galactic  latitudes  $\mathrm{|b|  >  5^{\circ}}$, several studies have shown that the double RC is seen only close to the minor axis of the bulge ($\mathrm{-4^{\circ} < l < +4^{\circ}}$ ). The magnitude separation between the two RCs increases with increasing Galactic latitude from 0.4 mag at $\mathrm{b=-5^{\circ}}$ to 0.7 mag at $\mathrm{b=-10^{\circ}}$ \citep{mcwilliam-zoccali+10, Saito+11, wegg-gerhard+13}.

\item At a fixed Galactic latitude, the fRC becomes gradually less populated towards lines of sight at increasing longitudes, until only the bRC can be detected. The opposite effect is observed towards negative Galactic longitudes \citep{mcwilliam-zoccali+10, Saito+11}.

\item Everywhere, the variation of the RC magnitude as well as the magnitude difference between the two RCs,
where there are two, is independent of the observational band.

\end{enumerate}

Figure~1 shows the luminosity function of a set of bulge fields located at a constant latitude strip at $\mathrm{b=-8^{\circ}}$ and for two additional fields along the minor axis to illustrate the points listed above. Each field is $1^\circ \times 1^\circ$ in size. To construct the luminosity functions we used the Vista Variables in the Via Lactea $\rm J$, $\rm K_{s}$ PSF photometric catalogues (Alonso-Garcia et al. 2015, in preparation) corrected for reddening using the BEAM calculator \citep{gonzalez+12}. 

\section{Multiple population spheroid and bar component scenario}

\citet{lee+15} presented a scenario where the double red  clump is the consequence of a classical spheroid
with two stellar populations, one of which  is significantly enhanced in helium. Specifically, a
  first population, called G1, has a  skewed metallicity distribution in the range $-0.75<$[Fe/H]$<+0.05$
  and a standard helium enrichment parameter, thus populating the faint RC.  A second, super-helium-rich
generation of stars (G2),  with $-0.55<$[Fe/H]$<+0.25$ and $Y=0.39$ populates  the bright RC.
\citet{lee+15} assumed that the global  metallicity distribution function peaks at [Fe/H]=$-0.1,$ but
  the two populations have  a mean metallicity difference of $\rm  \Delta[Fe/H]_{G2-G1}$=0.2 and ages of
12 Gyr for  G1 and 10 Gyr for  G2. In addition to  this classical spheroid component, they included a bar
population similar to  G1, but with an age of  10 Gyr and a narrow metallicity  distribution peaked at
  [Fe/H]=+0.15, with $\sigma_{\rm [Fe/H]}=0.1$.  They adopted a population ratio between  the bar and the
CB components of 2:1 at low latitudes, comparable to a population ratio at $\rm |b|\sim5.5^{\circ}$, and fully
dominated by the CB component at higher latitudes. \citet{lee+15} did not explicitly state what they refer
to as  low and high  latitudes, therefore we assumed for  our analysis  three  Galactic latitude strips  at $\rm
b=-3.0^{\circ}$, -$\rm5.5^{\circ}$, and $\rm -8.5^{\circ}$.

With this scenario, \citet{lee+15} claimed to be able to reproduce all  the key observations of the bulge. They found a
magnitude difference  of 0.5 mag between  the faint and bright RC, with negligible  differences in
colour. The  double RC would  not be  detected in the  metal-poor stars  since the  metallicity is
required  to  be  [Fe/H]>-0.4  for  the  super-helium-rich G2  to  form  and  remain  in  the  bRC.  The
latitude-longitude dependency of the double RC is then explained by the different contributions of
the  CB and  a bar  component with  a suitable  position angle  (cf. their  Fig.~4). The  RC of  the bar
population would be suitably placed between the faint and  bright RC of the CB component. Thus, since the
bar population would become less dominant at increasing  |b| and vanish at high latitudes, the
double RC  would become more prominent  and the magnitude difference  between fRC and bRC  would increase
with increasing latitude |b|.  On the other hand, at intermediate  latitudes ($\rm |b|=5.5^{\circ}$), the
effect of the bar position angle would cause the bulge RC to
overlap with one of the two RCs of the
spheoroid, so that the other, largely outnumbered, would not be detected. At positive longitude the bar is closer, hence its RC would overlap with the spheroid bRC, and vice versa at negative longitudes.

\section{Testing the scenario with the observed double red clump of the bulge}

In this  section  we  test the  multiple  population classical  bulge  (MCB)  scenario proposed  by
  \cite{lee+15} by comparing  it to some key observations  that, in our opinion, were  not properly taken
  into account in their paper. In particular, we focus on observations where the points (P1, P2, P3, and P4) listed
in Sect. 2 can be reliably tested. Figure~2 shows the bulge luminosity functions at Galactic longitudes
$\rm  l=-2^{\circ}$, $\rm  0^{\circ}$,  and  $\rm +2^{\circ}$  at  latitudes  $\rm b=-3.0^{\circ}$,  $\rm
-5.5^{\circ}$, and $\rm-8.5^{\circ}$.   For each line of sight, the underlying  red giant branch was
subtracted using an exponential function following the prescriptions of \citet{nataf+10}.

\begin{figure}[ht]
\centering
\includegraphics[width=9cm,angle=0]{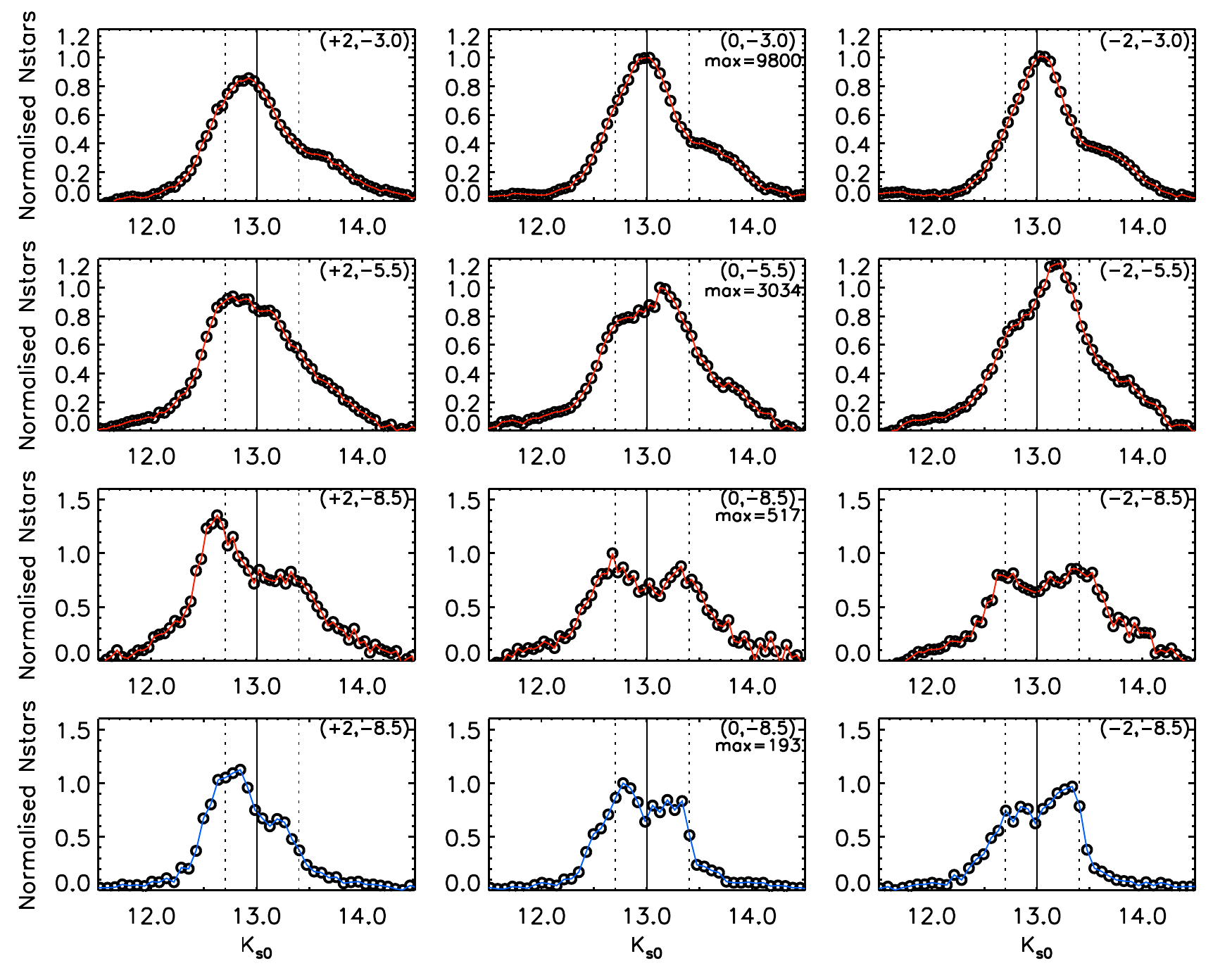}\\
\caption{RC magnitude distributions for lines of sight for Galactic longitudes l=-2$^{\circ}$, 0$^{\circ}$, and +2$^{\circ}$ and Galactic latitudes $\mathrm{b=-3.0^{\circ}}$, $\mathrm{-5.5^{\circ}}$, and $\mathrm{-8.5^{\circ}}$. The bottom panel shows the simulated magnitudes derived from the distance distribution of particles of the B/P bulge model from \citet{ness+14} at the corresponding lines of sight for $\mathrm{b=-8.5^{\circ}}$. All the distributions at the same Galactic latitude have been normalized by the maximum of the central distribution ($\mathrm{l=0^{\circ}}$). The solid black line shows the mean of the RC distribution at $\rm (l,b)=(0^{\circ},-3.0^{\circ})$ and the dashed lines show the maximum separation of the two RCs at $\rm (l,b)=(0^{\circ},-8.5^{\circ})$.}
\label{fig2}
\end{figure}

 In principle,  P1 can be satisfied within the MCB scenario of  \citet{lee+15} by placing  a population
 dominated by the bar in these inner regions. P4 can also be satisfied based on the negligible photometric bandpass dependency of the RC magnitude of G1 and G2 (S. Cassisi, private communication). We note, however, that the mean magnitude changes of the
 single RC, induced by the  bar position angle, should remain the same at all latitudes. The upper panels of Fig.~\ref{fig2}
show that the single  RC observed at low latitudes has a mean magnitude variation of $\rm \Delta  K_{s}=\pm0.2$ when moving from $\rm l=+2^\circ$ to  $\rm l=-2^\circ$. According to \citet{lee+15}, the bulge should be dominated by the bar at
these latitudes, therefore this is the maximum
 possible variation of the bar RC that is due to the bar position angle.  At higher $\rm{|b|}$,  the magnitude
 variation of  the bar RC invoked to erase the signature of the double RC should not exceed  0.2 mag.
   Furthermore, at higher latitudes ($\rm |b|=8.5^\circ$), no variations should be observed between
     the RC magnitude distribution at positive and negative longitudes, as this is  where the effect of
   the bar would be negligible, and we would only see the MCB.  In contrast, it is not possible to
     explain the  RC magnitude  variations observed at  $\rm b=-5.5^{\circ}$  as due  to the  bar RC
       shifts seen at $\rm b=-3.0^{\circ}$.  The observed changes in  the RC distributions  become even
     more dramatic  at $\rm b=-8.5^{\circ}$.  Here, not only is the separation between the fRC and bRC
     already 0.7 mag at $\rm l=0^{\circ}$, as expected  from P2, but the bRC  at $\rm l=+2^{\circ}$
     also becomes even more dominant with respect to the fRC, although no bar should be present here in the scenario of \citet{lee+15}. A Galactic latitude of $\rm b=-8.5^{\circ}$ corresponds to a height of  $\rm \sim1.2$ kpc from the plane for a population located at 8 kpc from the Sun.  Certainly,  for the bar population to be present at such large heights from the plane, a buckling instability process would be required. Similarly, the fact that the fRC (bRC) becomes less populated with increasing (decreasing) longitude, to the point where only one of the RCs is visible as described in P3 (Fig.~\ref{fig1}), firmly argues against the MCB scenario. The only way to explain P3 within the scenario of \citet{lee+15} would be to impose a rather unrealistic spatial distribution for populations G1 and G2 where only one of them is present at each side of the MCB.

On the other hand, the B/P nature of the bulge is able to naturally explain the observed changes in
the RC seen in Figs.~\ref{fig1} and~\ref{fig2}. To judge the plausibility of such a scenario, it is important to
  emphasize  that the double RC  was  discovered and  later  mapped by  \citet{Saito+11} using  2MASS
photometry. As a result of the high incompleteness of  2MASS in the innermost regions ($\rm |b|<3^{\circ}$), early
studies did not properly normalize the star  counts in the inner and outer  bulge. Because of
this, the first (qualitative) estimates of the density contrast of the X-shape were overestimated.
When VVV photometry became available, \citet{wegg-gerhard+13}  performed a 3D reconstruction of the shape
of the  whole bulge, using  a correct density  normalization.  From their results  it was clear  that the
contrast of the X-shaped structure  in the outer bulge is very similar to  that of external galaxies with
B/P bulges,  of which  several examples  are available in  the nearby  universe \citep[][]{laurikainen+15}. In fact, bars seen edge-on have a boxy shape, and most of them also have a peanut shape, which is due to the natural instability of bars that produces
some bending and buckling of the bar. It is then important to realize that the X-shape claimed to be present in the Milky Way is not a particularly exotic hypothesis, but a rather common  feature in barred galaxies.

To  support the latter statement, we compare here the  observed RC magnitude distributions with the output of a B/P bulge simulation from \citet{ness+14}. We note that we did not fine-tune the simulation, which  was not specifically made to  reproduce the Milky Way properties.   We use  it here  only because it  has a  pure B/P  bulge and  thus its
line-of-sight distance  distributions provide  the ideal  test for interpreting  the RC  observations. We
scaled the model size to be comparable to the B/P of  the Milky Way using a factor of 1.2 as described in
\citet{ness+14}.  We transformed the coordinates  of the model to place the Sun at  8 kpc from the centre
of the galaxy and the  bar to have a position angle of $\rm 27^{\circ}$  with respect to the Sun-Galactic
centre line-of-sight \citep{rattenbury+07, wegg-gerhard+13}. We converted the line-of-sight distance to every star particle  of the model to an observed  magnitude by  adopting an  absolute magnitude  for  the RC  of $\rm  M(K_s) =  -1.55$ mag \citep{gonzalez+11c}.  The
magnitude distribution of stars towards each line of sight was then convolved with a Gaussian FWHM of 0.21
mag to account for the observational  effects of the bulge RC population \citep{gerhard+12}. The
resulting simulated RC magnitudes for $\rm b=-8.5^{\circ}$ are shown in the bottom panel of Fig.~2. The B/P model
nicely reproduces  the double RC shape and variations seen at  higher latitudes.  The fact that these
magnitude distributions  were constructed directly from  their spatial distribution confirms that the RC
magnitude closely follows the  shape of the Milky Way bulge structure. The resemblance with the observed
distribution at $\rm b=-8.5^{\circ}$ certainly confirms the B/P bulge origin for the double RC.

\section{Concluding remarks}

We have provided a list of the observational properties of the double RC of the Milky Way bulge that must be satisfied when attempting to interpret its origin. We showed that these basic properties disagree with the scenario proposed by \citet{lee+15}, where the double RC originates from a MCB with two generations of stars each of them with a different helium enrichment, but instead they are a natural consequence of the B/P shape of the Milky Way bulge. The main reasons that support our conclusion is that a B/P bulge, which is very common in external galaxies, is able to consistently explain the presence (and absence) of a double RC where it is actually observed without fine-tuning any parameters. The bar orientation angle that best reproduces the observations \citep[27$^{\circ}$ according to][]{wegg-gerhard+13} is fully compatible with previous measurements \citep[e.g.][]{rattenbury+07}. In contrast, the work of \citet{lee+15} is based on the search for a mechanism able to produce a double RC everywhere in the Galactic bulge. This obviously needs another, specifically tailored mechanism to erase the double RC everywhere except in a narrow range of longitudes in the outer bulge. Such a task requires fine-tuning both the iron and helium distribution of G1 and G2 and of a suitable bar orientation angle and density distribution. Yet, as we have demonstrated, it still cannot explain the disappearance of the double RC away from the minor axis (in longitude) in the outer bulge. Instead, we showed that the line-of-sight distance distribution of a simulated galaxy with a B/P bulge naturally follows the observed RC magnitude distribution of the Milky Way bulge at high distances from the plane. 

We do consider that including stellar population variations such
as were presented in \citet{lee+15} are important to determine the precision of the RC as a distance indicator. The effect of stellar populations is currently considered to be minimal in morphological studies of the Milky Way bulge. Providing a set of corrections can help improving our current map of the B/P bulge structure of the Milky Way even further.

%__________________________________________________________________

\begin{acknowledgements}
We thank the anonymous referee for useful comments. We warmly thank Andy McWilliam for his thoughtful comments and Santi Cassisi for useful discussions about the RC magnitude variation as a function of helium abundance. MZ acknowledges funding from the BASAL CATA through grant PFB-06, and the Chilean Ministry of Economy through ICM grant to the Millennium Institute of Astrophysics and the support by Proyecto Fondecyt Regular 1150345. VPD is supported by STFC Consolidated grant \# ST/J001341/1. JA-G acknowledges support by the FIC-R Fund, allocated to the project 30321072. The simulation used in this study was run at the High Performance Computer Facility of the University of Central Lancashire.
\end{acknowledgements}

\bibliographystyle{aa}
\small
\bibliography{mybiblio_rev_full}

\begin{thebibliography}{24}
\expandafter\ifx\csname natexlab\endcsname\relax\def\natexlab#1{#1}\fi

\bibitem[{{Athanassoula}(2005)}]{Ath05b}
{Athanassoula}, E. 2005, \mnras, 358, 1477

\bibitem[{{Babusiaux} \& {Gilmore}(2005)}]{babusiaux+05}
{Babusiaux}, C. \& {Gilmore}, G. 2005, \mnras, 358, 1309

\bibitem[{{Bournaud} {et~al.}(2009){Bournaud}, {Elmegreen}, \&
  {Martig}}]{bournaud+09}
{Bournaud}, F., {Elmegreen}, B.~G., \& {Martig}, M. 2009, \apjl, 707, L1

\bibitem[{{Brook} \& {Christensen}(2015)}]{brook-christensen+15}
{Brook}, A.~M. \& {Christensen}, C. 2015, Laurikainen E., Peletier R., Gadotti
  D. A., eds., in press

\bibitem[{{Brook} {et~al.}(2011){Brook}, {Governato}, {Ro{\v s}kar}, {Stinson},
  {Brooks}, {Wadsley}, {Quinn}, {Gibson}, {Snaith}, {Pilkington}, {House}, \&
  {Pontzen}}]{brook+11}
{Brook}, C.~B., {Governato}, F., {Ro{\v s}kar}, R., {et~al.} 2011, \mnras, 415,
  1051

\bibitem[{{Cao} {et~al.}(2013){Cao}, {Mao}, {Nataf}, {Rattenbury}, \&
  {Gould}}]{cao+13}
{Cao}, L., {Mao}, S., {Nataf}, D., {Rattenbury}, N.~J., \& {Gould}, A. 2013,
  \mnras, 434, 595

\bibitem[{{Combes} \& {Sanders}(1981)}]{ComSan81}
{Combes}, F. \& {Sanders}, R.~H. 1981, \aap, 96, 164

\bibitem[{{Dekel} \& {Burkert}(2014)}]{dekel-burkert+14}
{Dekel}, A. \& {Burkert}, A. 2014, \mnras, 438, 1870

\bibitem[{{Gerhard} \& {Martinez-Valpuesta}(2012)}]{gerhard+12}
{Gerhard}, O. \& {Martinez-Valpuesta}, I. 2012, \apjl, 744, L8

\bibitem[{{Gonzalez} {et~al.}(2011){Gonzalez}, {Rejkuba}, {Minniti}, {Zoccali},
  {Valenti}, \& {Saito}}]{gonzalez+11c}
{Gonzalez}, O.~A., {Rejkuba}, M., {Minniti}, D., {et~al.} 2011, \aap, 534, L14

\bibitem[{{Gonzalez} {et~al.}(2013){Gonzalez}, {Rejkuba}, {Zoccali}, {Valent},
  {Minniti}, \& {Tobar}}]{gonzalez+13}
{Gonzalez}, O.~A., {Rejkuba}, M., {Zoccali}, M., {et~al.} 2013, \aap, 552, A110

\bibitem[{{Gonzalez} {et~al.}(2012){Gonzalez}, {Rejkuba}, {Zoccali}, {Valenti},
  {Minniti}, {Schultheis}, {Tobar}, \& {Chen}}]{gonzalez+12}
{Gonzalez}, O.~A., {Rejkuba}, M., {Zoccali}, M., {et~al.} 2012, \aap, 543, A13

\bibitem[{{Immeli} {et~al.}(2004){Immeli}, {Samland}, {Gerhard}, \&
  {Westera}}]{ImmSamGer04}
{Immeli}, A., {Samland}, M., {Gerhard}, O., \& {Westera}, P. 2004, \aap, 413,
  547

\bibitem[{{Laurikainen} \& {Salo}(2015)}]{laurikainen+15}
{Laurikainen}, E. \& {Salo}, H. 2015, ArXiv:1505.00590

\bibitem[{{Lee} {et~al.}(2015){Lee}, {Joo}, \& {Chung}}]{lee+15}
{Lee}, Y.-W., {Joo}, S.-J., \& {Chung}, C. 2015, ArXiv:1508.05942

\bibitem[{{McWilliam} {et~al.}(2010){McWilliam}, {Fulbright}, \&
  {Rich}}]{mcwilliam-fulbright-rich+10}
{McWilliam}, A., {Fulbright}, J., \& {Rich}, R.~M. 2010, in IAU Symposium, Vol.
  265, IAU Symposium, ed. K.~{Cunha}, M.~{Spite}, \& B.~{Barbuy}, 279--284

\bibitem[{{McWilliam} \& {Zoccali}(2010)}]{mcwilliam-zoccali+10}
{McWilliam}, A. \& {Zoccali}, M. 2010, \apj, 724, 1491

\bibitem[{{Nataf} {et~al.}(2010){Nataf}, {Udalski}, {Gould}, {Fouqu{\'e}}, \&
  {Stanek}}]{nataf+10}
{Nataf}, D.~M., {Udalski}, A., {Gould}, A., {Fouqu{\'e}}, P., \& {Stanek},
  K.~Z. 2010, \apjl, 721, L28

\bibitem[{{Ness} {et~al.}(2014){Ness}, {Debattista}, {Bensby}, {Feltzing},
  {Ro{\v s}kar}, {Cole}, {Johnson}, \& {Freeman}}]{ness+14}
{Ness}, M., {Debattista}, V.~P., {Bensby}, T., {et~al.} 2014, \apjl, 787, L19

\bibitem[{{Rattenbury} {et~al.}(2007){Rattenbury}, {Mao}, {Sumi}, \&
  {Smith}}]{rattenbury+07}
{Rattenbury}, N.~J., {Mao}, S., {Sumi}, T., \& {Smith}, M.~C. 2007, \mnras,
  378, 1064

\bibitem[{{Saito} {et~al.}(2011){Saito}, {Zoccali}, {McWilliam}, {Minniti},
  {Gonzalez}, \& {Hill}}]{Saito+11}
{Saito}, R.~K., {Zoccali}, M., {McWilliam}, A., {et~al.} 2011, \aj, 142, 76

\bibitem[{{Stanek} \& {Garnavich}(1998)}]{stanek+98}
{Stanek}, K.~Z. \& {Garnavich}, P.~M. 1998, \apjl, 503, L131

\bibitem[{{Wegg} \& {Gerhard}(2013)}]{wegg-gerhard+13}
{Wegg}, C. \& {Gerhard}, O. 2013, \mnras, 435, 1874

\bibitem[{{Zoccali}(2010)}]{zoccali+10}
{Zoccali}, M. 2010, in IAU Symposium, Vol. 265, IAU Symposium, ed. K.~{Cunha},
  M.~{Spite}, \& B.~{Barbuy}, 271--278

\end{thebibliography}

\end{document}